\documentclass[11pt,fleqn,twoside]{article}
\usepackage{amsfonts,amssymb,latexsym,maple2e}
\makeatletter
\newcommand{\prava}{\footnotesize\it
\begin{flushright}
\begin{minipage}{18cm}%{6cm}%9.6
Copyright \copyright 1998 by R. Ab\l amowicz
\end{minipage}
\end{flushright}}

\newcommand{\name}[1]{\begin{flushleft}
                       \LARGE \bf #1
                       \end{flushleft}\vspace{-3mm}}

\newcommand{\Author}[1]{\begin{flushleft}
                       \it #1 \end{flushleft}}

\newcommand{\Adress}[1]{\begin{flushleft}
                       \it #1 \end{flushleft}}

\newcommand{\Date}[1]{\begin{flushleft}
                      \small  \it #1 \end{flushleft}}

\newcommand{\ehkol}{Author \ name}
\newcommand{\ohkol}{Article \ name}
\renewcommand{\@evenhead}{
\hspace*{-3pt}\raisebox{-15pt}[\headheight][0pt]{\vbox{\hbox to \textwidth
{\thepage \hfil \ehkol}\vskip4pt \hrule}}}
\renewcommand{\@oddhead}{
\hspace*{-3pt}\raisebox{-15pt}[\headheight][0pt]{\vbox{\hbox to \textwidth
{\ohkol \hfil \thepage}\vskip4pt\hrule}}}
\renewcommand{\@evenfoot}{}
\renewcommand{\@oddfoot}{}

\newcommand{\be}{\begin{equation}}
\newcommand{\ee}{\end{equation}}
\newcommand{\ba}{\hspace*{-5pt}\begin{array}}
\newcommand{\ea}{\end{array}}

\newcommand{\ds}{\displaystyle}
\makeatother

\newcommand{\cl}{C \kern -0.1em \ell}     %Clifford algebra
\newcommand{\MBS}{\raise 1pt \hbox{$\phantom{\scriptstyle>}$\space}}

\begin{document}
\setcounter{page}{294}
\thispagestyle{empty}

\renewcommand{\ehkol}{R. Ab\l amowicz}
\renewcommand{\ohkol}{Matrix Exponential via Clif\/ford Algebras}

\begin{flushleft}
\footnotesize \sf
Journal of Nonlinear Mathematical Physics \qquad 1998, V.5, N~3,
\pageref{ablamowicz-fp}--\pageref{ablamowicz-lp}.
\hfill {\sc Article}
\end{flushleft}

\vspace{-5mm}

{\renewcommand{\footnoterule}{}
{\renewcommand{\thefootnote}{}
 \footnote{\prava}}

\name{Matrix Exponential via Clif\/ford Algebras}\label{ablamowicz-fp}

\Author{Rafa\l \  AB\L AMOWICZ}

%\thanks{Expanded version of a talk presented at the Special Session
%on `Octonions and Clifford Algebras', 1997 Spring Western Sectional
%921st Meeting of the American Mathematical Society, Oregon State
%University, Corvallis, OR, 19 - 20 April 1997.}

\Adress{$^*$~Department of Mathematics and Physics, Gannon University,
Erie, PA 16541\\
E-mail: ablamowicz@gannon.edu
\renewcommand{\thefootnote}{}\footnote{$^*$
Address after July 1, 1998: Department of Mathematics, Box 5054,
Tennessee Technological University, Cookeville, TN 38505, E-mail:
rablamowicz@tntech.edu
}
}

\Date{Received March 17, 1998; Accepted May 15, 1998}

\begin{flushright}
\begin{minipage}{10cm}
\small
Expanded version of a talk presented at the Special Session
on `Octonions and Clif\/ford Algebras', 1997 Spring Western Sectional
921st Meeting of the American Mathematical Society, Oregon State
University, Corvallis, OR, 19--20 April 1997.
\end{minipage}
\end{flushright}

\begin{abstract}
\noindent
We use isomorphism $\varphi$ between matrix algebras and simple
orthogonal Clif\/ford algebras $\cl(Q)$ to compute matrix exponential
${\mathrm e}^{A}$ of a real, complex, and quaternionic matrix $A.$ The
isomorphic image $p=\varphi(A)$ in $\cl(Q),$ where the quadratic form
$Q$ has a suitable signature $(p,q),$ is exponentiated modulo a
minimal polynomial of $p$ using Clif\/ford exponential. Elements of
$\cl(Q)$ are treated as symbolic multivariate polynomials in
Grassmann monomials. Computations in $\cl(Q)$ are performed with a
Maple package {\sc `CLIFFORD'}.  Three examples of matrix
exponentiation are given.
%{\bf Keywords:} Clif\/ford algebra, primitive idempotent, spinor
%representation, Grassmann algebra, quaternions, matrix norm.
\end{abstract}

\section{Introduction}

Exponentiation of a numeric $n \times n$ matrix $A$ is needed when
solving a system of dif\/ferential equations ${\bf x}' = A{\bf x}$,
${\bf x}(0)={\bf x}_0,$ in order to represent its solution in a form
${\mathrm e}^{At}{\bf x}_0.$ It is well known that the exponential form of the
solution remains valid when $A$ is
not diagonalizable, provided the following def\/inition of ${\mathrm e}^{A}$ is
adopted:
\begin{equation}
{\mathrm e}^A=\sum_{k=0}^{\infty} \frac{A^k}{k!}, \qquad \mbox {where} \quad
A^0=I.
\label{ablamowicz:eq:1}
\end{equation}
Equation~(\ref{ablamowicz:eq:1}) means that the sequence of partial sums
$\ds S_n\equiv\sum\limits_{k=0}^{n} A^k/k! \rightarrow {\mathrm e}^A$ entrywise.
Equivalently, (\ref{ablamowicz:eq:1}) implies that $\| S_n-{\mathrm e}^A \|_1
\rightarrow 0$ where $\| A \|_1$ denotes matrix $1$-norm def\/ined as
the maximum of $\{\| A_j \|_1,\,j=1,\ldots,n\},$ $A_j$ is the $j$th
column of a $A,$ and $\| A_j \|_1$ is the
$1$-vector norm on ${\mathbb C}^n$ def\/ined as $\ds \| {\bf x} \|_1 =
\sum\limits_{i=1}^{n}
|x_i|.$ However, for several reasons, there is no obvious
way\footnote[1]{It is possible to compute the exponential ${\mathrm e}^{At}$
with a help of the Laplace transform method applied
to an appropriate
system of dif\/ferential equations \cite{ablamowicz:McDonald}.} to implement
def\/inition (\ref{ablamowicz:eq:1}) on a computer, unless of course $A$ is
diagonalizable, that is, when $A$ has a complete set of linearly
independent eigenvectors (cf. \cite{ablamowicz:Scheick97}).

Another approach to solving ${\bf x}' = A{\bf x}$ is to f\/ind Jordan canonical
form $J$ of the matrix $A.$ Let $P$ be a nonsingular matrix such that
$P^{-1}AP=J.$ Then, if a change of basis is made such that ${\bf x} =
P{\bf y},$ the matrix equation ${\bf x}' = A{\bf x}$ is transformed
into ${\bf y}' =
J{\bf y}$ and, at least theoretically, its solution is represented as
${\mathrm e}^{Jt}{\bf c}$ for some constant vector ${\bf c}.$
However, since the
Jordan form is extremely discontinuous on a set of all $n \times n$
matrices, numeric computations of $J$ are seriously ill-posed (cf.
\cite{ablamowicz:Scheick97,ablamowicz:KwakHong}).

In this paper we present another approach to exponentiate a matrix,
let it be numeric or symbolic, with real, complex, or quaternionic
entries, totally dif\/ferent from the linear algebra methods.  It
relies on the well-known isomorphism between matrix algebras over
${\mathbb R}$, ${\mathbb C}$,  or ${\mathbb H}$,
 and simple orthogonal Clif\/ford algebras
(cf. \cite{ablamowicz:Crum90,ablamowicz:AblLoun95,ablamowicz:AblLounParra96,ablamowicz:Loun81}).  This is not
a matrix method in the sense that elements of the real Clif\/ford
algebra $\cl(Q)$ are not viewed here as matrices but instead they are
treated as symbolic multivariate polynomials in some basis Grassmann
monomials. This is
possible due to the linear isomorphism $\cl(V,Q) \simeq \bigwedge V.$ The
critical exponentiation is done in the real Clif\/ford algebra
$\cl_{p,q}$ over $Q$ with a suitable signature $(p,q)$ depending
whether the given matrix $A$ has real, complex, or quaternionic
entries.  Three examples of computation of the matrix exponential
with a Maple package {\sc `CLIFFORD'} (cf.
\cite{ablamowicz:Abl96,ablamowicz:Abl97,ablamowicz:MapleV4}) are presented below. The Reader is encouraged to
repeat these computations.
}

In order to f\/ind matrix exponential ${\mathrm e}^A,$ the following steps
will be taken:

\begin{itemize}
\item[--] We will view elements of $\cl_{p,q}$ as real multivariate
polynomials in basis Grassmann or Clif\/ford monomials.

\item[--] We will f\/ind explicit spinor (left-regular) representation
$\gamma$ of $\cl_{p,q}$ in a minimal left ideal $S=\cl_{p,q}f$ generated
by a primitive idempotent $f.$

\item[--] For a matrix $A$ (numeric or symbolic) in the matrix ring
${\mathbb R}(n),$ ${\mathbb C}(n)$ or ${\mathbb H}(n)$ where
$n=2^{m-1}$, $m=\left[\frac12(p+q)\right],$ we will f\/ind its
isomorphic image $p=\varphi(A)$ in $\cl_{p,q}.$\footnote[2]{The brackets
$\lbrack\,\cdot\,\rbrack$ denote the f\/loor function}

\item[--] We will f\/ind a {\it real} minimal polynomial $p(x)$ of $p$
and then a formal power series $\exp(p) \bmod p(x)$ in $\cl_{p,q}.$

\item[--] We will check the truncation error of the power series $\exp(p)$
in $\cl_{p,q}$ via a polynomial norm, or in a matrix norm, both built
into Maple.\footnote[3]{It is also possible to use the ${\mathbb R}^{n'}$
topology where $n' = 2^n$, $n=p+q.$}

\item[--] We will map $\exp(p)$ back to the matrix ring ${\mathbb R}(n),$
${\mathbb C}(n)$  or ${\mathbb H}(n)$ to get~$\exp(A).$
\end{itemize}

Before we proceed, let's recall certain useful facts about orthogonal
Clif\/ford algebras $\cl_{p,q}.$ For more information see~\cite{ablamowicz:Crum90}.

\begin{itemize}
\item[--] If $p-q\not=1\bmod 4$ then $\cl_{p,q}$ is a simple algebra of
dimension $2^n$, $n=p+q,$ isomorphic with a full matrix algebra with
entries in ${\mathbb R}$, ${\mathbb C},$ or ${\mathbb H}.$

\item[--] If $p-q=1\bmod 4$ then $\cl_{p,q}$ is a semi-simple algebra of
dimension $2^n$, $n=p+q,$ containing two copies of a full matrix
algebra with entries in ${\mathbb R}$ or ${\mathbb H}$
projected out by two central idempotents
$\frac12(1\pm{\bf e}_1{\bf e}_2\cdots{\bf e}_n).$\footnote[4]{For the
purpose of this paper, it is enough to consider simple Clif\/ford algebras
only.}

\item[--] $\cl_{p,q}$ has a faithful representation as a matrix
algebra with entries in ${\mathbb R},$ ${\mathbb C},$
${\mathbb H}$ or
${\mathbb R}\oplus {\mathbb R},$ ${\mathbb H} \oplus
{\mathbb H}$ depending whether $\cl_{p,q}$ is simple
 or semisimple.

\item[--] Any primitive idempotent $f$ in $\cl_{p,q}$ is expressible as a
product
\begin{equation}
f = \frac12 (1\pm e_{T_1})\frac12 (1\pm e_{T_2})\cdots\frac12 (1\pm e_{T_k})
\label{ablamowicz:eq:finclpq}
\end{equation}
where $\{e_{T_1},e_{T_2},\ldots,e_{T_k}\}$, $k=q-r_{q-p},$ is a set of
commuting basis monomials with square $1,$ and $r_i$ is the
Radon-Hurwitz number def\/ined by the recursion $r_{i+8}=r_{i}+4$ and

\vspace{-5mm}

\begin{center}
\tabcolsep=10pt
\begin{tabular}{c|cccccccc}
\setlength{\unitlength}{10mm}
 $i$  & $0$ & $1$ & $2$ & $3$ & $4$ & $5$ & $6$ & $7$ \\ \hline
$r_i$ & $0$ & $1$ & $2$ & $2$ & $3$ & $3$ & $3$ & $3$
\end{tabular} \,.
\end{center}

\item[--] $\cl_{p,q}$ has a complete set of $2^k$ primitive
idempotents each with $k$ factors as in (\ref{ablamowicz:eq:finclpq}).

\item[--] The division ring ${\mathbb K}=f \cl_{p,q} f$ is isomorphic to
${\mathbb R}$
or ${\mathbb C}$ or ${\mathbb H}$ when $(p-q) \bmod 8$ is $0,1,2,$ or $3,7$
or $4,5,6.$

\item[--] The mapping $S \times {\mathbb K} \rightarrow S,$ or
$(\psi,\lambda) \rightarrow \psi\lambda$ def\/ines a right ${\mathbb K}$-linear
structure on the spinor space $S=\cl_{p,q}f$ (cf. \cite{ablamowicz:Loun81}).

\end{itemize}

\noindent
{\bf Example 1.}  In $\cl_{3,1} \simeq {\mathbb R}(4)$ we have $k=2$ and
$f=\frac12 (1+{\bf e}_1) \frac12 (1+{\bf e}_{34}),$
${\bf e}_{34}={\bf e}_3 {\bf e}_4 = {\bf e}_3
\wedge {\bf e}_4$ is a primitive idempotent. The ring ${\mathbb K}
 \simeq {\mathbb R}$ is just
spanned by $\{1\}_{\mathbb R}$ and a real basis for $S=\cl_{3,1}f$ may be
generated by $\{ 1,{\bf e}_2,{\bf e}_3,{\bf e}_{23}\}_{\mathbb R}$ (here
${\bf e}_{23} = {\bf e}_2 {\bf e}_3 = {\bf e}_2 \wedge  {\bf e}_3.)$

\smallskip

\noindent
{\bf Example 2.} In $\cl_{3,0} \simeq {\mathbb C}(2)$ we have $k=1$ and
$f=\frac12 (1+{\bf e}_1)$
is a primitive idempotent. The ring ${\mathbb K} \simeq {\mathbb C}$
may be spanned
by $\{1,{\bf e}_{23} \}_{\mathbb R}$ and a basis for $S=\cl_{3,0}f$ over
${\mathbb K}$ may be generated by $\{1,{\bf e}_2\}_{\mathbb K}.$

\smallskip

\noindent
{\bf Example 3.}
In $\cl_{1,3} \simeq {\mathbb H}(2),$ the Clif\/ford polynomial $ f=\frac12
(1+{\bf e}_{14}),$  ${\bf e}_{14} = {\bf e}_1 {\bf e}_4 =
{\bf e}_1 \wedge {\bf e}_4,$ is a primitive idempotent.
Thus, the ring ${\mathbb K} \simeq {\mathbb H}$ may be spanned by $\{
1,{\bf e}_2,{\bf e}_3,{\bf e}_{23}\}_{\mathbb R}$ and a basis for
$S=\cl_{1,3}f$ as a right-quaternionic space over ${\mathbb K}$ may be
generated by $\{1,{\bf e}_1\}_{\mathbb K}.$

\section{Exponential of a real matrix}

We now proceed to exponentiate a real $4 \times 4$ matrix using the
spinor representation $\gamma$ of $\cl_{3,1}$ from Example~1. Instead
of $\cl_{3,1}$ one could also use $\cl_{2,2},$ the Clif\/ford algebra
of the neutral signature $(2,2),$ since $\cl_{2,2} \simeq {\mathbb
R}(4).$ From now on ${\bf e}_{ij}={\bf e}_i{\bf e}_j=
{\bf e}_i \wedge {\bf e}_j,\, i \neq j,$ ${\mathbb K} = \{
Id \}_{{\mathbb R}} \simeq {\mathbb R},$ and $Id$ denotes the unit element of
$\cl_{3,1}$ in {\sc `CLIFFORD'}

Recall the following facts about the simple algebra $\cl_{3,1} \simeq
{\mathbb R}(4)$ and its spinor space~$S$:
\begin{itemize}
\item[--]$\cl_{3,1}=\lbrace 1,{\bf e}_i,\,{\bf e}_{ij},\,{\bf
e}_{ijk},\, {\bf e}_{ijkl}
\rbrace_{\mathbb R},\quad i < j < k < l,\  i,j,k,l = 1,\ldots,4.$

\item[--]$S=\cl_{3,1}f=\lbrace
f_1=f,\,f_2={\bf e}_{2}f,\,f_3={\bf e}_{3}f,\,f_4={\bf e}_{23}f
\rbrace_{\mathbb K}.$

\item[--]Each basis monomial ${\bf e}_{ijkl}$ has a unique matrix
$\gamma_{{\bf e}_{ijkl}}$ representation in the spinor basis
$f_i,\,i=1,\ldots,4.$ For example, the basis $1$-vectors
${\bf e}_1,{\bf e}_2,{\bf e}_3,{\bf e}_4 $ are represented under
$\gamma$ as:
\end{itemize}
\be
\ba{cc}
\gamma_{{\bf e}_1} = \pmatrix{
1 &  0 & 0 & 0  \cr
0 & -1 & 0 & 0  \cr
0 & 0 & -1 & 0  \cr
0 & 0 & 0 & 1},
&
\gamma_{{\bf e}_2} = \pmatrix{
0 & 1 & 0 & 0 \cr
1 & 0 & 0 & 0 \cr
0 & 0 & 0 & 1 \cr
0 & 0 & 1 & 0}
\\ & \\
\gamma_{{\bf e}_3} = \pmatrix{
0 & 0 & 1 &  0 \cr
0 & 0 & 0 & -1 \cr
1 & 0 & 0 &  0 \cr
0 & -1 & 0 & 0},
&
\gamma_{{\bf e}_4} = \pmatrix{
0 & 0 & -1 & 0 \cr
0 & 0 &  0 & 1 \cr
1 & 0 &  0 & 0 \cr
0 & -1 & 0 & 0}.
\end{array}
\label{ablamowicz:eq:gammasincl31}
\ee
Since $\gamma: {\mathbb R}(4) \rightarrow \cl_{3,1}$ is a linear
isomorphism of algebras, matrices representing Clif\/ford monomials of
higher ranks
are matrix products of matrices shown in (\ref{ablamowicz:eq:gammasincl31}). For
example, $\gamma_{{\bf e}_{ijkl}} = \gamma_{{\bf e}_i}
\gamma_{{\bf e}_j} \gamma_{{\bf e}_k} \gamma_{{\bf e}_l}$:
\be
\gamma_{{\bf e}_{1234}} = \gamma_{{\bf e}_1} \gamma_{\bf e_2}
\gamma_{{\bf e}_3} \gamma_{{\bf e}_4} =
\pmatrix{
0 & 1 & 0 & 0  \cr
-1 & 0 & 0 & 0 \cr
0 & 0 & 0 & 1  \cr
0 & 0 & -1 & 0}.
\label{ablamowicz:eq:3}
\ee
Then, a matrix representing any Clif\/ford polynomial may be found by
the linearity of $\gamma.$

Relevant information about $\cl_{3,1}$ is stored in {\sc `CLIFFORD'} and
can be retrieved as follows:
\begin{maplegroup}
\begin{mapleinput}
\mapleinline{active}{1d}{
   restart:with(Cliff3):dim:=4:B:=linalg[diag](1,1,1,-1):}{%}
\mapleinline{active}{1d}{eval(makealiases(dim)):data:=clidata();}{%}
\end{mapleinput}
\end{maplegroup}
\begin{maplegroup}
\mapleresult
\begin{maplelatex}
\begin{eqnarray*}
\lefteqn{data :=} \\
 & & [real, \,4, \,simple, \,
cmulQ({\displaystyle \frac {1}{2}} \,Id +
            {\displaystyle \frac {1}{2}} \, e1, \,
            {\displaystyle \frac {1}{2}} \, Id +
            {\displaystyle \frac {1}{2}} \, e34), \\
&&             [Id, \,e2, \,e3, \,e23], [Id], \,[Id, \,e2, \,e3, \,e23]]
\end{eqnarray*}
\end{maplelatex}
\end{maplegroup}
\noindent
In the Maple list {\it data} above,

\begin{itemize}
\item[--] {\it real,} $4,$ and {\it simple} mean that $\cl_{3,1}$ is a
simple algebra isomorphic to~${\mathbb R}(4).$

\item[--] The fourth element {\tt data[4]} in the list {\tt 'data'} is
a primitive idempotent $f$ written as a Clif\/ford product of two
Clif\/ford polynomials (Clif\/ford product in orthogonal Clif\/ford
algebras is realized in {\sc `CLIFFORD'} through a procedure {\tt
'cmulQ'}).

\item[--] The list $[Id,\,e2,\,e3,\,e23]$ contains generators of the
spinor space $S=\cl_{3,1}f$ over the reals ${\mathbb R}$ (compare with
Example 1 above).

\item[--] The list $[Id\,]$ contains the only basis element of the
f\/ield ${\mathbb K} \subset \cl_{3,1},$ that is, the identity element of
$\cl_{3,1}.$

\item[--] The f\/inal list $[Id,\,e2,\,e3,\,e23]$ contains generators of the
spinor space $S=\cl_{3,1}f$ over the f\/ield ${\mathbb K}.$ In this case it
coincides with {\tt data[5]} since ${\mathbb K} \simeq {\mathbb R}.$
\end{itemize}

\noindent
Thus, a real spinor basis in $S$ consists of the following four
polynomials:
\begin{maplegroup}
\begin{mapleinput}
\mapleinline{active}{1d}{f1:=f;f2:=cmulQ(e2,f);f3:=cmulQ(e3,f);f4:=
cmulQ(e23,f);}{%}
\end{mapleinput}
\end{maplegroup}
\be
\ba{ll}
f1 \mbox{\rm :=} {\displaystyle \frac {1}{4}} \,Id + {\displaystyle
\frac {1}{4}} \,e34 +
      {\displaystyle \frac {1}{4}} \,e1 + {\displaystyle \frac {1}{4}} \,e134,
&
f2 \mbox{\rm :=} {\displaystyle \frac {1}{4}} \,e2 + {\displaystyle
\frac {1}{4}} \,e234 -
{\displaystyle \frac {1}{4}} \,e12 - {\displaystyle \frac {1}{4}}
\,e1234
\\[2.8ex]
f3 \mbox{\rm :=} {\displaystyle \frac {1}{4}} \,e3 + {\displaystyle
\frac {1}{4}} \,e4 -
{\displaystyle \frac {1}{4}} \,e13 - {\displaystyle \frac {1}{4}}
\,e14,
&
f4 \mbox{\rm :=} {\displaystyle \frac {1}{4}} \,e23 + {\displaystyle
\frac {1}{4}} \,e24 +
{\displaystyle \frac {1}{4}} \,e123 + {\displaystyle \frac {1}{4}}
\,e124
\ea
\label{ablamowicz:eq:fsincl31}
\ee

Procedure {\tt 'matKrepr'} allows us now to compute $16$ matrices $m[i]$
representing each basis monomial in~$\cl_{3,1}.$
\begin{mapleinput}
\mapleinline{active}{1d}{for i from 1 to 16 do}{%}
\mapleinline{active}{1d}{lprint (`The basis element`,clibas[i],\newline
\MBS  \qquad \qquad `is represented by the following matrix:);}{%}
\mapleinline{active}{1d}{m[i]:=subs(Id=1,matKrepr(clibas[i])) od:}{%}
\end{mapleinput}

Let's def\/ine a $4 \times 4$ real matrix $A$ without a complete set of
eigenvectors. Therefore, $A$ cannot be diagonalized.
\begin{maplegroup}
\begin{mapleinput}
\mapleinline{active}{1d}
   {A:=linalg[matrix](4,4,[0,1,0,0,-1,2,0,0,-1,1,1,0,-1,1,0,1]);}{%}
\mapleinline{active}{1d}{linalg[eigenvects](A);#A has incomplete set
of eigenvectors}{%}
\end{mapleinput}
\mapleresult
\begin{maplelatex}
\[
A :=  \left[
{\begin{array}{rrrr}
0 & 1 & 0 & 0 \\
-1 & 2 & 0 & 0 \\
-1 & 1 & 1 & 0 \\
-1 & 1 & 0 & 1
\end{array}}
 \right]
\]
\end{maplelatex}
\end{maplegroup}
\vspace{-5mm}

\be
[1, \,4, \,\{[0, \,0, \,1, \,0], \,[1, \,1, \,0, \,0], \,[0, \,0,
\,0, \,1]\}]
\label{ablamowicz:eq:eigen1}
\ee
Maple output in (\ref{ablamowicz:eq:eigen1}) shows that $A$ has only one eigenvalue
$\lambda=1$ with an algebraic multiplicity $4$ and a geometric
multiplicity $3.$

In the Appendix, one can f\/ind a procedure {\tt 'phi'} which gives the
isomorphism $\varphi$ from ${\mathbb R}(4)$ to $\cl_{3,1}.$ It can f\/ind the
image $p= {\mbox {\tt phi}}(A)$ of any real $4 \times 4$ matrix $A$
using the
 previously computed matrices $m[i].$ In particular, the image $p$ of $A$
 under $\varphi$ is computed as follows:
\begin{maplegroup}
\begin{mapleinput}
\mapleinline{active}{1d}{FBgens:=[Id]; #assigning a basis element of K}{%}
\mapleinline{active}{1d}{p:=phi(A,m,FBgens); #finding the image of A
in Cl(3,1)}{%}
\end{mapleinput}
\end{maplegroup}
\begin{equation}
p \mbox{\rm :=} Id - {\displaystyle \frac {1}{2}} \,e1 - {\displaystyle
 \frac {1}{2}} \,e3 -
{\displaystyle \frac {1}{2}} \,e4 + {\displaystyle \frac {1}{2}}
\,e12 -
{\displaystyle \frac {1}{2}} \,e23 - {\displaystyle \frac {1}{2}}
\,e24 -
{\displaystyle \frac {1}{2}} \,e134 + {\displaystyle \frac {1}{2}}
\,e1234
\label{ablamowicz:eq:pincl31}
\ee

Let's go back to the exponentiation problem.  So far we have found a
Clif\/ford polynomial $p$ in $\cl_{3,1}$ which is the isomorphic image
of $A.$ We will now compute a sequence of f\/inite power series
expansions of $p$
up to a specif\/ied order $N.$ Procedure {\tt 'sexp'} (def\/ined in the
Appendix) f\/inds these expansions, which are just Clif\/ford
polynomials, modulo
the minimal polynomial $p(x)$ of $p.$  The minimal polynomial $p(x)$ can
be computed using a procedure
{\tt 'climinpoly'}.
\begin{maplegroup}
\begin{mapleinput}
\mapleinline{active}{1d}{p(x)=climinpoly(p);}{%}
\end{mapleinput}
\end{maplegroup}
\begin{equation}
p(x)=x^{2} - 2\,x + 1
\label{ablamowicz:eq:polincl31}
\end{equation}
It can be easily verif\/ied that the polynomial (\ref{ablamowicz:eq:polincl31}) is
satisf\/ied by $p=\varphi(A)$ and that it is also the minimal
polynomial of $A.$
\begin{maplegroup}
\begin{mapleinput}
\mapleinline{active}{1d}{cmul(p,p)-2*p+Id; #p satisfies its own
minimal polynomial}{%}
\end{mapleinput}
\mapleresult
\begin{maplelatex}
\[
0
\]
\end{maplelatex}
\end{maplegroup}
\begin{maplegroup}
\begin{mapleinput}
\mapleinline{active}{1d}{
   linalg[minpoly](A,x); #matrix A has the same minimal polynomial as p}{%}
\end{mapleinput}
\mapleresult
\begin{maplelatex}
\[
x^{2} - 2\,x + 1
\]
\end{maplelatex}
\end{maplegroup}
\noindent
A f\/inite sequence of say $20$ Clif\/ford polynomials approximating
$\exp(p)$ can now be computed.
\begin{maplegroup}
\begin{mapleinput}
\mapleinline{active}{1d}{
   N:=20:for i from 1 to N do p.i:=sexp(p,i) od:# we want 20 polynomials}{%}
\end{mapleinput}
\end{maplegroup}
For example, Maple displays polynomial $p_{20}$ as follows:
\begin{maplegroup}
\begin{mapleinput}
\mapleinline{active}{1d}{p\_lim:=p.20;}{}
\end{mapleinput}
\mapleresult
\begin{maplelatex}
\begin{eqnarray*}
\lefteqn{p\_lim := {\displaystyle \frac
{6613313319248080001}{2432902008176640000}} \,Id -
{\displaystyle \frac {82666416490601}{60822550204416}} \,e1
- {\displaystyle \frac {82666416490601}{60822550204416}} \,e3}
\\[1.0ex]
& & \mbox{} - {\displaystyle \frac {82666416490601}{60822550204416}}
\,e4 +
{\displaystyle \frac {82666416490601}{60822550204416}} \,e12 -
{\displaystyle \frac {82666416490601}{60822550204416}} \,e23
\\[1.0ex]
& & \mbox{} - {\displaystyle \frac {82666416490601}{60822550204416}}
\,e24 -
{\displaystyle \frac {82666416490601}{60822550204416}} \,e134 +
{\displaystyle \frac {82666416490601}{60822550204416}} \,e1234
\mbox{\hspace{30pt}}
\end{eqnarray*}
\end{maplelatex}
\end{maplegroup}
\noindent
Having computed the approximation polynomials
$p_1,p_2,\ldots,p_N,\,N=20,$ one can show that the sequence converges
to some limiting polynomial
$p_{lim}$ by verifying that $|p_i - p_j| < \epsilon$ for $i,j > M,\,
M$ suf\/f\/iciently large, in one of the Maple's built-in polynomial
norms.

Finally, we map back $p_{lim}$ into a $4 \times 4$ matrix which
approximates $\exp (A)$ up to and including the terms of order~$N.$
\begin{maplegroup}
\begin{mapleinput}
\mapleinline{active}{1d}{expA:=0:for i from 1 to nops(clibas) do}{%}
\mapleinline{active}{1d}{\qquad expA:=evalm(expA+coeff(p_{lim},
clibas[i])*m[i])od:}{%}
\mapleinline{active}{1d}{evalm(expA); #the matrix exponent of A}{%}
\end{mapleinput}
\mapleresult
\begin{maplelatex}
\[
 \left[
{\ba{cccc}
\frac {1}{2432902008176640000} &
\frac {82666416490601}{30411275102208} & 0 & 0 \\ [2ex]
\frac {-82666416490601}{30411275102208} &
\frac {7775794614048301}{1430277488640000} & 0 & 0 \\ [2ex]
\frac {-82666416490601}{30411275102208} &
\frac {82666416490601}{30411275102208} &
\frac {6613313319248080001}{2432902008176640000} & 0 \\ [2ex]
\frac {-82666416490601}{30411275102208} &
\frac {82666416490601}{30411275102208} & 0 &
\frac {6613313319248080001}{2432902008176640000}
\end{array}}
 \right]
\]
\end{maplelatex}
\end{maplegroup}
\noindent
Although $A$ had an incomplete set of eigenvectors, Maple can f\/ind
$\exp(A)$ in a closed form.
\begin{maplegroup}
\begin{mapleinput}
\mapleinline{active}{1d}{mA:=linalg[exponential](A);  }{%}
\end{mapleinput}
\mapleresult
\begin{maplelatex}
\[
mA :=  \left[ {\begin{array}{cccc}
0 & e & 0 & 0 \\
 - e & 2\,e & 0 & 0 \\
 - e & e & e & 0 \\
 - e & e & 0 & e
\end{array}}
 \right]
\]
\end{maplelatex}
\end{maplegroup}
\noindent
Notice that our result is very close to the Maple closed-form result:
\begin{maplegroup}
\begin{mapleinput}
\mapleinline{active}{1d}{map(evalf,evalm(expA));}{%}
\end{mapleinput}
\mapleresult
\begin{maplelatex}
\begin{eqnarray*}
\lefteqn{[.41103176233121648585\,10^{-18}\,, \,
2.7182818284590452349\,, \,0\,, \,0]} \\
 & & \!\!\![-2.7182818284590452349\,, \,5.4365636569180904703\,, \,0\,
, \,0] \\
 & & \!\!\![-2.7182818284590452349\,, \,2.7182818284590452349\,, \,
2.7182818284590452353\,, \,0] \\
 & & \!\!\![-2.7182818284590452349\,, \,2.7182818284590452349\,, \,0\,
, \,2.7182818284590452353]
\end{eqnarray*}
\end{maplelatex}
\end{maplegroup}
The $1$-norm of the dif\/ference matrix between $mA$ and $expA$ can be
computed in Maple as follows:
\begin{maplegroup}
\begin{mapleinput}
\mapleinline{active}{1d}{evalf(linalg[norm](mA-expA,1));}{%}
\end{mapleinput}
\mapleresult
\begin{maplelatex}
\[
.2\,10^{-17}
\]
\end{maplelatex}
\end{maplegroup}

\vspace{-6mm}

\section{Exponential of a complex matrix}
In this section we exponentiate a complex $2 \times 2$ matrix using a
spinor representation of $\cl_{3,0} \simeq {\mathbb C}(2)$ (see  Example~2
above).
Note that instead of using $\cl_{3,0},$ one could also use
$\cl_{1,2}$ since $\cl_{1,2} \simeq {\mathbb C}(2).$ As before,
${\bf e}_{ijk}={\bf e}_i {\bf e}_j {\bf e}_k = {\bf e}_i \wedge {\bf
e}_j \wedge  {\bf e}_k ,$
$i,j,k=1,\ldots,3,$ ${\mathbb K} = \{ Id,\,{\bf e}_{23
}\}_{\mathbb R} \simeq {\mathbb C},$ ${\bf e}_{23}^2=-Id,$
where $Id$ denotes the unit element of $\cl_{3,0}$ in {\sc `CLIFFORD'}.

Recall these facts about the simple algebra $\cl_{3,0}$ and its
spinor space $S$:
\begin{itemize}
\item[--] $\cl_{3,0}=\lbrace 1,\,{\bf e}_i,\,{\bf e}_{ij},\,{\bf e}_{ijk}
\rbrace_{\mathbb R},\ i < j < k.$

\item[--] $S=\cl_{3,0}f=\lbrace
f_1=f,\,f_2={\bf e}_{2}f,\,f_3={\bf e}_{3}f,\,f_4={\bf e}_{23}f
\rbrace_{\mathbb R}.$

\item[--] $S=\cl_{3,0}f=\lbrace f_1=f,\,f_2={\bf e}_{2}f
\rbrace_{\mathbb K}.$
\end{itemize}
For example, the basis $1$-vectors are represented in the spinor
basis $\{f_1,f_2\}$ by these three matrices in ${\mathbb K}(2)$ well known as
the {\it Pauli matrices}:
\be
\gamma_{{\bf e}_1} =  \pmatrix{
1 & 0 \cr
0 & -1},
\qquad
\gamma_{{\bf e}_2} = \pmatrix{ 0 & 1 \cr
1 & 0},
\qquad
\gamma_{{\bf e}_3} = \pmatrix{
0 & -{\bf e}_{23} \cr
{\bf e}_{23} & 0}.
\label{ablamowicz:eq:4}
\ee
The following information about $\cl_{3,0}$ is stored in {\sc `CLIFFORD'}:
\begin{maplegroup}
\begin{mapleinput}
\mapleinline{active}{1d}{dim:=3:B:=linalg[diag](1,1,1):}{%}
\mapleinline{active}{1d}{data:=clidata();}{%}
\end{mapleinput}
\mapleresult
\begin{maplelatex}
\[
data := [complex, \,2, \,simple, \,
{\displaystyle \frac {1}{2}} \,Id + {\displaystyle \frac {1}{2}} \,e1, \,
[Id,\, e2,\, e3,\, e23], \,
[Id,\, e23], \, [Id,\, e2]]
\]
\end{maplelatex}
\end{maplegroup}

Now we def\/ine a Grassmann basis in $\cl_{3,0},$ assign a primitive
idempotent to $f,$ and generate a spinor basis for $S=\cl_{3,0}f.$
\begin{maplegroup}
\begin{mapleinput}
\mapleinline{active}{1d}{clibas:=cbasis(dim); #ordered basis in Cl(3,0)}{%}
\end{mapleinput}
\mapleresult
\begin{maplelatex}
\[
clibas := [Id, \,e1, \,e2, \,e3, \, e12, \,e13, \,e23, \,e123]
\]
\end{maplelatex}
\end{maplegroup}
\begin{maplegroup}
\begin{mapleinput}
\mapleinline{active}{1d}{f:=data[4];  #a primitive idempotent in Cl(3,0)}{%}
\end{mapleinput}
\mapleresult
\begin{maplelatex}
\[
f := {\displaystyle \frac {1}{2}} \,Id + {\displaystyle \frac {1}{2}} \,e1
\]
\end{maplelatex}
\end{maplegroup}
\begin{maplegroup}
\begin{mapleinput}
\mapleinline{active}{1d}{sbasis:=minimalideal(clibas,f,'left'); #find
a real basis in Cl(B)f}{%}
\end{mapleinput}
\mapleresult
\begin{maplelatex}
\begin{eqnarray*}
\lefteqn{sbasis := }\\
& & \!\!\![[{\displaystyle \frac {1}{2}} \,Id +
{\displaystyle \frac {1}{2}} \,e1, \,{\displaystyle \frac {1}{2}} \,e2 -
{\displaystyle \frac {1}{2}} \,e12, \,{\displaystyle \frac {1}{2}} \,e3 -
 {\displaystyle \frac {1}{2}} \,e13, \,{\displaystyle \frac {1}{2}} \,e23
+
{\displaystyle \frac {1}{2}} \,e123], \,[Id, \,e2, \,e3, \,e23], \, left]
 \end{eqnarray*}
\end{maplelatex}
\end{maplegroup}
\begin{maplegroup}
\begin{mapleinput}
\mapleinline{active}{1d}{fbasis:=Kfield(sbasis,f); #find a basis for
the field K}{%}
\end{mapleinput}
\mapleresult
\begin{maplelatex}
\[
fbasis := [[{\displaystyle \frac {1}{2}} \,Id +
{\displaystyle \frac {1}{2}} \,e1, \,{\displaystyle \frac {1}{2}} \,e23 +
 {\displaystyle \frac {1}{2}} \,e123], \,[Id, \,e23]]
\]
\end{maplelatex}
\end{maplegroup}
\begin{maplegroup}
\begin{mapleinput}
\mapleinline{active}{1d}{SBgens:=sbasis[2];#generators for a real basis
 in S }{%}
\end{mapleinput}
\mapleresult
\begin{maplelatex}
\[
SBgens := [Id, \,e2, \,e3, \,e23]
\]
\end{maplelatex}
\end{maplegroup}
\begin{maplegroup}
\begin{mapleinput}
\mapleinline{active}{1d}{FBgens:=fbasis[2]; #generators for K}{%}
\end{mapleinput}
\mapleresult
\begin{maplelatex}
\[
FBgens := [Id, \,e23]
\]
\end{maplelatex}
\end{maplegroup}

In the above, {\tt 'sbasis'} is a ${\mathbb K}$-basis returned for
$S=\cl_{3,0}f.$ Since in the current signature $(3,0)$ we have
${\mathbb K}=\{ Id,\,e23
\}_{{\mathbb R}} \simeq {\mathbb C},$ $\mbox{\tt cmulQ}(e23,\,e23) =
-Id,$ and $\cl_{3, 0} \simeq {\mathbb C}(2),$ the
output from {\tt 'spinorKbasis'} shown below
has two basis vectors and their generators modulo $f$:
\begin{maplegroup}
\begin{mapleinput}
\mapleinline{active}{1d}{Kbasis:=spinorKbasis(SBgens,f,FBgens,'left');}{%}
\end{mapleinput}
\mapleresult
\begin{maplelatex}
\[
Kbasis := [[{\displaystyle \frac {1}{2}} \,Id + {\displaystyle \frac
{1}{2}} \,e1, \,
            {\displaystyle \frac {1}{2}} \,e2 - {\displaystyle \frac {1}{
2}} \,e12], \,
            [Id, \,e2], \,left]
\]
\end{maplelatex}
\end{maplegroup}
\begin{maplegroup}
\begin{mapleinput}
\mapleinline{active}{1d}{cmulQ(f,f); #verifying that f is an idempotent}{%}
\end{mapleinput}
\mapleresult
\begin{maplelatex}
\[
{\displaystyle \frac {1}{2}} \,Id + {\displaystyle \frac {1}{2}} \,e1
\]
\end{maplelatex}
\end{maplegroup}
\noindent
Note that the second list in {\tt 'Kbasis'} contains generators of
the f\/irst list modulo the idempotent $f.$ Thus, the spinor basis in
$S$ over ${\mathbb K}$ consists of the following two polynomials:
\begin{maplegroup}
\begin{mapleinput}
\mapleinline{active}{1d}{for i from 1 to nops(Kbasis[1]) do
f.i:=Kbasis[1][i] od;}{%}
\end{mapleinput}
\end{maplegroup}
\begin{equation}
f1 := {\displaystyle \frac {1}{2}} \,Id + {\displaystyle \frac
{1}{2}} \,e1,
\qquad
f2 := {\displaystyle \frac {1}{2}} \,e2 - {\displaystyle \frac {1}{2}}
\,e12
\label{ablamowicz:eq:fsincl30}
\end{equation}
We are in a position now to compute matrices $m[i]$ representing
basis elements in $\cl_{3,0}.$ We will only display Clif\/ford-algebra
valued matrices representing the $1$-vectors $\{{\bf e}_1,{\bf e}_2,
{\bf e}_3\}$ and the unit pseudoscalar ${\bf e}_{123}={\bf e}_1{\bf
e}_2{\bf e}_3.$

\begin{maplegroup}
\begin{mapleinput}
\mapleinline{active}{1d}{for i from 1 to nops(clibas) do}{%}
\mapleinline{active}{1d}{lprint (`The basis element`,clibas[i],\newline
\MBS  \qquad \qquad `is represented by the following matrix:`);}{%}
\mapleinline{active}{1d}{m[i]:=subs(Id=1,matKrepr(clibas[i])) od:}{%}
\end{mapleinput}

\mapleresult
\begin{maplettyout}
The basis element   e1   is represented by the following matrix:
\end{maplettyout}

\begin{maplelatex}
\[
{m_{2}} :=  \left[
{\begin{array}{rr}
1 & 0 \\
0 & -1
\end{array}}
 \right]
\]
\end{maplelatex}

\begin{maplettyout}
The basis element   e2   is represented by the following matrix:
\end{maplettyout}

\begin{maplelatex}
\[
{m_{3}} :=  \left[
{\begin{array}{rr}
0 & 1 \\
1 & 0
\end{array}}
 \right]
\]
\end{maplelatex}

\begin{maplettyout}
The basis element   e3   is represented by the following matrix:
\end{maplettyout}

\begin{maplelatex}
\[
{m_{4}} :=  \left[
{\begin{array}{cc}
0 &  - e23 \\
e23 & 0
\end{array}}
 \right]
\]
\end{maplelatex}

\begin{maplettyout}
The basis element   e123   is represented by the following matrix:
\end{maplettyout}

\begin{maplelatex}
\[
{m_{8}} :=  \left[
{\begin{array}{cc}
e23 & 0 \\
0 & e23
\end{array}}
 \right]
\]
\end{maplelatex}

\end{maplegroup}

As an example, let's def\/ine a complex $2 \times 2$ matrix $A$ and
let's f\/ind its eigenvectors:
\begin{maplegroup}
\begin{mapleinput}
\mapleinline{active}{1d}{A:=linalg[matrix](2,2,[1+2*I,1-3*I,1-I,-2*I]);
 #defining A}{%}
\mapleinline{active}{1d}{linalg[eigenvects](A);}{%}
\end{mapleinput}
\mapleresult
\begin{maplelatex}
\[
A :=  \left[
{\begin{array}{cc}
1 + 2\,I & 1 - 3\,I \\
1 - I &  - 2\,I
\end{array}}
 \right]
\]
\end{maplelatex}
\vspace{-5mm}

\begin{maplelatex}
\begin{eqnarray*}
\lefteqn{[{\displaystyle \frac {1}{2}}  + {\displaystyle \frac {1
}{2}} \,\sqrt{ - 23 - 8\,I}, \,1, \,\{ \left[  \!  -
{\displaystyle \frac {3}{4}}  + {\displaystyle \frac {1}{4}} \,
\sqrt{ - 23 - 8\,I} + I + {\displaystyle \frac {1}{2}} \,I\,(
{\displaystyle \frac {1}{2}}  + {\displaystyle \frac {1}{2}} \,
\sqrt{ - 23 - 8\,I}), \,1 \!  \right] \}],}  \\
 & & \![{\displaystyle \frac {1}{2}}  - {\displaystyle \frac {1}{2}
} \,\sqrt{ - 23 - 8\,I}, \,1, \,\{ \left[  \!  - {\displaystyle
\frac {3}{4}}  - {\displaystyle \frac {1}{4}} \,\sqrt{ - 23 - 8\,
I} + I + {\displaystyle \frac {1}{2}} \,I\,({\displaystyle
\frac {1}{2}}  - {\displaystyle \frac {1}{2}} \,\sqrt{ - 23 - 8\,
I}), \,1 \!  \right] \}]
\end{eqnarray*}
\end{maplelatex}
\end{maplegroup}

The image of $A$ in $\cl_{3,0}$ under the isomorphism $\varphi:
{\mathbb C}(2) \rightarrow \cl_{3,0}$ can now be computed. Recall that {\tt
'FBgens'} def\/ined above contained the basis elements of the complex
f\/ield ${\mathbb K}$ in $\cl_{3,0}.$
\begin{maplegroup}
\begin{mapleinput}
\mapleinline{active}{1d}{evalm(A);p:=phi(A,m,FBgens); #finding image
of A in Cl(3,0)}{%}
\end{mapleinput}
\mapleresult
\begin{maplelatex}
\[
 \left[
{\begin{array}{cc}
1 + 2\,I & 1 - 3\,I \\
1 - I &  - 2\,I
\end{array}}
 \right]
\]
\end{maplelatex}
\vspace{-5mm}

\begin{maplelatex}
\[
p := {\displaystyle \frac {1}{2}} \,Id + {\displaystyle
\frac {1}{2}} \,e1 + e2 + e3 + 2\,e13 + 2\,e23
\]
\end{maplelatex}
\end{maplegroup}

Thus, we have found a Clif\/ford polynomial $p$ in $\cl_{3,0}$ which is the
isomorphic image of $A.$ We will now compute a sequence of f\/inite
power expansions of $p$ up to and including power $N=30$ using the
procedure {\tt 'sexp'}.  This sequence of Clif\/ford polynomials should
converge to a polynomial $p_{lim},$ the image under $\varphi$ of the
matrix exponential $\exp(A).$ First, we f\/ind the {\it real\/} minimal
polynomial $p(x)$ of $p$ (called {\tt 'pol'} in Maple).
\begin{maplegroup}
\begin{mapleinput}
\mapleinline{active}{1d}{pol:=climinpoly(p); #find the real minimal
polynomial of p}{%}
\end{mapleinput}
\mapleresult
\begin{maplelatex}
\[
pol := x^{4} - 2\,x^{3} + 13\,x^{2} - 12\,x + 40
\]
\end{maplelatex}
\end{maplegroup}
\begin{maplegroup}
\begin{mapleinput}
\mapleinline{active}{1d}{&c(p$4)-2*&c(p$3)+13*&c(p$2)-12*p+40*Id;#checking
that p satisfies pol}{%}
\end{mapleinput}
\mapleresult
\begin{maplelatex}
\[
0
\]
\end{maplelatex}
\end{maplegroup}
\noindent
Observe that matrix $A$ has the following {\it complex\/} minimal
polynomial {\tt 'pol2'}:
\begin{maplegroup}
\begin{mapleinput}
\mapleinline{active}{1d}{pol2:=linalg[minpoly](A,x);}{%}
\end{mapleinput}
\mapleresult
\begin{maplelatex}
\[
pol2 := 6 + 2\,I - x + x^{2}
\]
\end{maplelatex}
\end{maplegroup}
\begin{maplegroup}
\begin{mapleinput}
\mapleinline{active}{1d}{evalm(&*(A$2)-A+6+2*I);}{%}
\end{mapleinput}
\mapleresult
\begin{maplelatex}
\[
 \left[
{\begin{array}{rr}
0 & 0 \\
0 & 0
\end{array}}
 \right]
\]
\end{maplelatex}
\end{maplegroup}
Furthermore, since $\{ Id, e123\}_{\mathbb R}$ is another copy of the complex
f\/ield ${\mathbb K}$ in $\cl_{3,0},$ we can easily verify that the Clif\/ford
polynomial $p$ also satisf\/ies the complex minimal polynomial {\tt
'pol2'} of $A$ if we replace $1$ with $Id$ and $I$ with $e123,$ namely:
\begin{maplegroup}
\begin{mapleinput}
\mapleinline{active}{1d}{&c(p$2)-p+6*Id+2*e123;}{%}
\end{mapleinput}
\mapleresult
\begin{maplelatex}
\[
0
\]
\end{maplelatex}
\end{maplegroup}
\noindent
On the other hand, matrix $A$ of course satisf\/ies the polynomial {\tt 'pol'}:
\begin{maplegroup}
\begin{mapleinput}
\mapleinline{active}{1d}{evalm(&*(A$4)-2*&*(A$3)+13*&*(A$2)-12*A+40);}{%}
\end{mapleinput}
\mapleresult
\begin{maplelatex}
\[
 \left[
{\begin{array}{rr}
0 & 0 \\
0 & 0
\end{array}}
 \right]
\]
\end{maplelatex}
\end{maplegroup}
\noindent
As expected, the complex minimal polynomial of $A$ is a factor of the
real minimal polynomial of~$p$:
\begin{maplegroup}
\begin{mapleinput}
\mapleinline{active}{1d}{divide(pol,pol2);}{%}
\end{mapleinput}
\mapleresult
\begin{maplelatex}
\[
true
\]
\end{maplelatex}
\end{maplegroup}
\begin{maplegroup}
\begin{mapleinput}
\mapleinline{active}{1d}{pol3:=quo(pol,pol2,x);}{%}
\end{mapleinput}
\mapleresult
\begin{maplelatex}
\[
pol3 := x^{2} - x + 6 - 2\,I
\]
\end{maplelatex}
\end{maplegroup}
Let's check that $\mbox{{\tt pol3}} \ast \mbox{{\tt pol2}} =
\mbox{{\tt pol}}$:
\begin{maplegroup}
\begin{mapleinput}
\mapleinline{active}{1d}{pol;expand(pol3 * pol2);}{%}
\end{mapleinput}
\mapleresult
\begin{maplelatex}
\[
x^{4} - 2\,x^{3} + 13\,x^{2} - 12\,x + 40
\]
%\end{maplelatex}
%\begin{maplelatex}
\[
x^{4} - 2\,x^{3} + 13\,x^{2} - 12\,x + 40
\]
\end{maplelatex}
\end{maplegroup}

The following loop computes Clif\/ford polynomials $p_i$ approximating
$\exp(p)$ in $\cl_{3,0}.$ We will only display polynomial $p_{30}$
and assign it to $p_{lim}.$
\begin{maplegroup}
\begin{mapleinput}
\mapleinline{active}{1d}{Digits:=20:}{%}
\mapleinline{active}{1d}{N:=30:for i from 1 to N do p.i:=sexp(p,i) od;}{%}
\mapleinline{active}{1d}{p_lim:=p.N:}{%}
\end{mapleinput}
\mapleresult
\begin{maplelatex}
\begin{eqnarray*}
\lefteqn{\hspace*{-26pt}p30 :=\! - {\displaystyle \frac
{739418826545208898275600203389}{544108430383981658741145600000}}
\,Id\! +\! {\displaystyle \frac
{140606618686769098555631609225939}{176835239874794039090872320000000}}
\,e1} \\[1.0ex]
& & \hspace*{-12mm} - {\displaystyle \frac
{13294860446171527820401106221093}{88417619937397019545436160000000}}
\,e2 + {\displaystyle \frac
{5429376085448859186420447465893}{12631088562485288506490880000000}}
\,e3 \\[1.0ex]
& & \hspace*{-12mm} + {\displaystyle \frac
{50830755859220399836279191881837}{44208809968698509772718080000000}}
\,e13 + {\displaystyle \frac
{15796535483801410769637551225479}{22104404984349254886359040000000}}
\,e23 \\[1.0ex]
& &  \hspace*{-12mm}- {\displaystyle \frac
{537129223345642211370021843709}{1184164552732995797483520000000}}
\,e123
- {\displaystyle \frac
{24569201649575451209456052913}{84691206836587183472640000000}} \,e12
\end{eqnarray*}
\end{maplelatex}
\end{maplegroup}
By picking up numeric coef\/f\/icients of the basis monomials in the
subsequent approximations to $\exp(p),$ one can get an idea about the
approximation errors.
\begin{maplegroup}
\begin{mapleinput}
\mapleinline{active}{1d}{sort([op(L:=cliterms(p_lim))],bygrade):}{%}
\mapleinline{active}{1d}{for i from 1 to nops(L) do}{%}
\mapleinline{active}{1d}{\qquad L.i:=map(evalf,[seq(coeff(p.j,L[i]),j=
1..N)]) od:}{%}
\mapleinline{active}{1d}{approxerror:=\newline
\MBS  \qquad \qquad max(seq(min(seq(abs(L.j[i]-L.j[i-1]),
i=2..N)), j=1..nops(L)));}{%}
\end{mapleinput}
\mapleresult
\begin{maplelatex}
\[
approxerror := .1\,10^{-19}
\]
\end{maplelatex}
\end{maplegroup}

Having computed the f\/inite sequence of polynomials $p_i$ one can
again show by using Maple's built-in polynomial norm functions that
this is a convergent sequence. For example, in the inf\/inity norm one
gets $|p_{29}-p_{30}| < .6 \times 10^{-20}$ and $|p_i-p_j|
\rightarrow 0$ as $i,j \rightarrow \infty.$

Thus, we have found an approximation $p_{lim}$ to the power series
expansion of $\exp(p)$ in $\cl_{3,0}$ up to and including terms of
degree $N=30.$ Finally, we map back $p_{lim}$ into a $2 \times 2$
complex matrix which approximates $\exp (A).$ We expand $p_{lim}$
over the matrices $m[i]$:
\begin{maplegroup}
\begin{mapleinput}
\mapleinline{active}{1d}{expA:=0: for i from 1 to nops(clibas) do}{%}
\mapleinline{active}{1d}{\qquad expA:=evalm(expA+coeff(p_lim,clibas[i]
)*m[i]) od:}{%}
\mapleinline{active}{1d}{evalm(expA); #the matrix exponent of A}{%}
\end{mapleinput}
\mapleresult
\begin{maplelatex}
\begin{eqnarray*}
\lefteqn{\left[ {\vrule height0.80em width0em depth0.80em} \right. \! \!
- {\displaystyle \frac
{7121749995744556670281318348249}{12631088562485288506490880000000}}
+
{\displaystyle \frac
{596909308415457533523842428577}{2286662584587853953761280000000}}
\,e23,} \\
 & &  \hspace*{-15.24pt}
- {\displaystyle \frac
{7789021393665659776614645092453}{17683523987479403909087232000000}}
- {\displaystyle \frac
{6228189267183180110479443301}{3942814712927403324211200000}} \,e23
   \! \!\left. {\vrule height0.80em width0em depth0.80em} \right]  \\
 & &  \hspace*{-15.24pt}
\lefteqn{ \left[ {\vrule height0.80em width0em depth0.80em} \right. \!  \!
{\displaystyle \frac
{12355386075985243242271013020079}{88417619937397019545436160000000}}
-
{\displaystyle \frac
{340405770696784948489921131029}{472821496991427912007680000000}}
\,e23,} \\
 & &  \hspace*{-15.24pt}
- {\displaystyle \frac
{31743144776163499207933472943947}{14736269989566169924239360000000}}
- {\displaystyle \frac
{5959141766058476626587221301857}{5101016534849828050698240000000}}
\,e23 \! \! \left. {\vrule height0.80em width0em depth0.80em} \right]
\end{eqnarray*}
\end{maplelatex}
\end{maplegroup}

Maple can f\/ind the exponent of $A$ in a closed form with its {\tt
'linalg[exponential]'} command.  We won't display the result but we
will just compare it numerically with our result saved in {\tt 'expA'}.
\begin{maplegroup}
\begin{mapleinput}
\mapleinline{active}{1d}{mA:=linalg[exponential](A):}{%}
\end{mapleinput}
\end{maplegroup}
Let's replace the monomial $e2we3$ in {\tt 'expA'} with the imaginary
unit $I$ used by Maple and let's apply {\tt 'evalf'} to the entries
of {\tt 'expA'}:
\begin{maplegroup}
\begin{mapleinput}
\mapleinline{active}{1d}{fexpA:=subs(e2we3=I,map(evalf,evalm(expA)));
}{%}
\end{mapleinput}
\mapleresult
\begin{maplelatex}
\begin{eqnarray*}
\lefteqn{fexpA :=} \\
 & &  [ - .56382709696901085353 + .26103952215715461164\,I\,,  \\
& & \mbox{\hspace{10pt}} - .44046771442052942162 -
1.5796302186766888057\,I] \\
 & &  [.13973895796712599250 - .71994563035478140661\,I\,,  \\
& & \mbox{\hspace{10pt}} - 2.1540827359052753813 -
1.1682263182928324795\,I]
\end{eqnarray*}
\end{maplelatex}
\end{maplegroup}
\begin{maplegroup}
\begin{mapleinput}
\mapleinline{active}{1d}{fmA:=map(evalf,mA); #applying 'evalf' to mA}{%}
\end{mapleinput}
\mapleresult
\begin{maplelatex}
\begin{eqnarray*}
\lefteqn{fmA :=} \\
 & &  [ - .56382709696901085362 + .26103952215715461158\,I\,,  \\
& & \mbox{\hspace{10pt}} - .44046771442052942180 -
1.5796302186766888058\,I] \\
 & &  [.13973895796712599243 - .71994563035478140663\,I\,,  \\
& & \mbox{\hspace{10pt}} - 2.1540827359052753816 -
1.1682263182928324795\,I]
\end{eqnarray*}
\end{maplelatex}
\end{maplegroup}
Let's check the $1$-norm of the dif\/ference matrix between {\tt 'fmA'}
and {\tt 'fexpA'}:
\begin{maplegroup}
\begin{mapleinput}
\mapleinline{active}{1d}{evalf(linalg[norm](fmA-fexpA,1));}{%}
\end{mapleinput}
\mapleresult
\begin{maplelatex}
\[
.5059126028\,10^{-18}
\]
\end{maplelatex}
\end{maplegroup}
The f\/loating-point approximation {\tt 'fexpA'} to $\exp(A)$ is within
approximately $.5 \times 10^{-18}$ in the matrix $\| \cdot \|_1$ norm
to the closed matrix exponential computed by Maple.

\section{Exponential of a quaternionic matrix}

In order to exponentiate a quaternionic $2 \times 2$ matrix, we will use
the spinor representation of $\cl_{1,3} \simeq {\mathbb H}(2)$ (see Example 3
above). Note that two other algebras could be used instead of
$\cl_{1,3},$ namely,
$\cl_{0,4}$ and $\cl_{4,0}$ since both are isomorphic to ${\mathbb H}(2).$ As
before ${\bf e}_{ij}={\bf e}_i {\bf e}_j = {\bf e}_i \wedge {\bf
e}_j,$ $i,j=1,\ldots,4,$ but
this time ${\mathbb K} = \{ Id,{\bf e}_2,{\bf e}_3,{\bf e}_{23}
\}_{\mathbb R} \simeq {\mathbb H}.$

Recall the following facts about the simple algebra $\cl_{1,3}$ and
its spinor space $S$:
\begin{itemize}
\item[--] $\cl_{1,3}=\lbrace 1,{\bf e}_i,{\bf e}_{ij},{\bf
e}_{ijk},{\bf e}_{ijkl}
\rbrace_{\mathbb R}, \ i<j<k<l.$

\item[--] $S=\cl_{1,3}f=\lbrace
f_1=f,f_2={\bf e}_{2}f,f_3={\bf e}_{3}f,f_4={\bf e}_{23}f
\rbrace_{\mathbb R}.$

\item[--] $S=\cl_{1,3}f=\lbrace f_1=f,f_2={\bf e}_{1}f
\rbrace_{\mathbb K}.$
\end{itemize}
For example, the basis $1$-vectors ${\bf e}_1,{\bf e}_2,{\bf e}_3,
{\bf e}_4$ are represented~by:
\be
\gamma_{{\bf e}_1} =  \pmatrix{
0 & 1 \cr
1 & 0},
\
\gamma_{{\bf e}_2} = \pmatrix{
{\bf e}_2 & 0 \cr
0 & -{\bf e}_2},
\
\gamma_{{\bf e}_3} = \pmatrix{
{\bf e}_3 & 0 \cr
0 & -{\bf e}_3},
\
\gamma_{{\bf e}_4} = \pmatrix{
0 & -1 \cr
1 & 0}.
\label{ablamowicz:eq:5}
\ee
In order to compute the spinor representation of $\cl_{1,3},$ we proceed
as follows:
\begin{maplegroup}
\begin{mapleinput}
\mapleinline{active}{1d}{data:=clidata(linalg[diag](1,-1,-1,-1));}{%}
\end{mapleinput}
\mapleresult
\begin{maplelatex}
\begin{eqnarray*}
\lefteqn{data := [quaternionic, \,2, \,simple,\,
{\displaystyle \frac {1}{2}} \,Id + {\displaystyle \frac {1}{2}} \,e14, \,
[Id, \,e1, \,e2, \,e3, \,e12, \,e13, \,e23, \,e123], } \\
 & & [Id, \,e2, \,e3, \,e23], \,[Id, \,e1]] %\mbox{\hspace{272pt}}
\end{eqnarray*}
\end{maplelatex}
\end{maplegroup}
We def\/ine a Grassmann basis in $\cl_{1,3},$ assign a primitive idempotent
 to $f,$ and generate a spinor basis for $S=\cl_{1,3}f.$
\begin{maplegroup}
\begin{mapleinput}
\mapleinline{active}{1d}{clibas:=cbasis(dim); #ordered basis in Cl(1,3)}{%}
\end{mapleinput}
\mapleresult
\begin{maplelatex}
\begin{eqnarray*}
\lefteqn{clibas :=}\\
& & [Id,\,e1,\,e2,\,e3,\,e4,\,e12,\,e13,\,e14,\,e23,\,e24,\,e34,
\,e123,\,e124,\,e134,\,e234,\,e1234]
\end{eqnarray*}
\end{maplelatex}
\end{maplegroup}
\begin{maplegroup}
\begin{mapleinput}
\mapleinline{active}{1d}{f:=data[4];  #a primitive idempotent in Cl(1,3)}{%}
\end{mapleinput}
\mapleresult
\begin{maplelatex}
\[
f := {\displaystyle \frac {1}{2}} \,Id + {\displaystyle \frac {1}{2}} \,e14
\]
\end{maplelatex}
\end{maplegroup}
Next, we compute a real basis in the spinor space $S=\cl_{1,3}f$ using
the command {\tt 'minimalideal'}:
\begin{maplegroup}
\begin{mapleinput}
\mapleinline{active}{1d}{sbasis:=minimalideal(clibas,f,'left');#find a
real basis in Cl(B)f}{%}
\end{mapleinput}
\mapleresult
\begin{maplelatex}
\begin{eqnarray*}
\lefteqn{sbasis := [[{\displaystyle \frac {1}{2}} \,Id + {\displaystyle
 \frac {1}{2}} \,e14, \,
{\displaystyle \frac {1}{2}} \,e1 + {\displaystyle \frac {1}{2}}
\,e4, \,
{\displaystyle \frac {1}{2}} \,e2 - {\displaystyle \frac {1}{2}}
\,e124, \,
{\displaystyle \frac {1}{2}} \,e3 - {\displaystyle \frac {1}{2}}
\,e134, \, {\displaystyle \frac {1}{2}} \,e12 - {\displaystyle
\frac {1}{2}} \,e24, } \\
& & \mbox{} {\displaystyle \frac {1}{2}} \,e13 - {\displaystyle \frac
{1}{2}} \,e34, \,
{\displaystyle \frac {1}{2}} \,e23 + {\displaystyle \frac {1}{2}}
\,e1234, \,
{\displaystyle \frac {1}{2}} \,e123 + {\displaystyle \frac {1}{2}}
\,e234], \, \\
& & \mbox{} [Id,\,e1,\,e2,\,e3,\,e12,\,e13,\,e23,\,e123],\, left]
%\mbox{\hspace{150pt}}
\end{eqnarray*}
\end{maplelatex}
\end{maplegroup}
In the following, we compute a basis for the subalgebra ${\mathbb K}$:
\begin{maplegroup}
\begin{mapleinput}
\mapleinline{active}{1d}{fbasis:=Kfield(sbasis,f); #a basis for the
field K}{%}
\end{mapleinput}
\mapleresult
\begin{maplelatex}
\begin{eqnarray*}
\lefteqn{fbasis :=}\\
& &  [[{\displaystyle \frac {1}{2}} \,Id + {\displaystyle \frac
{1}{2}} \,e14, \,
{\displaystyle \frac {1}{2}} \,e2 - {\displaystyle \frac {1}{2}}
\,e124, \,
{\displaystyle \frac {1}{2}} \,e3 - {\displaystyle \frac {1}{2}}
\,e134, \,
{\displaystyle \frac {1}{2}} \,e23 + {\displaystyle \frac {1}{2}}
\,e1234], \, [Id, \,e2, \,e3, \,e23]]
\end{eqnarray*}
\end{maplelatex}
\end{maplegroup}
\begin{maplegroup}
\begin{mapleinput}
\mapleinline{active}{1d}{SBgens:=sbasis[2];#generators for a real
basis in S }{%}
\end{mapleinput}
\mapleresult
\begin{maplelatex}
\[
SBgens := [Id, \,e1, \,e2, \,e3, \, e12, \,e13, \,e23, \,e123]
\]
\end{maplelatex}
\end{maplegroup}
Thus, a possible set of generators for ${\mathbb K}$ is:
\begin{maplegroup}
\begin{mapleinput}
\mapleinline{active}{1d}{FBgens:=fbasis[2]; #generators for K}{%}
\end{mapleinput}
\end{maplegroup}
\begin{equation}
FBgens := [Id, \,e2, \,e3, \,e23]
\label{ablamowicz:eq:FBgensincl30}
\end{equation}

In the above, {\tt 'sbasis'} is a real basis for $S=\cl_{1,3}f.$
Since in the current signature $(1,3)$ we have that ${\mathbb K} = \{
Id,\,e2,\,e3,\,e23\}_{\mathbb R} \simeq {\mathbb H}$ and
$\cl_{1,3} = {\mathbb H}(2),$ the
output from {\tt 'spinorKbasis'} shown below has two basis vectors
and their generators modulo $f$ for $S$ over ${\mathbb K}$:
\begin{maplegroup}
\begin{mapleinput}
\mapleinline{active}{1d}{Kbasis:=spinorKbasis(SBgens,f,FBgens,'left'); }{%}
\end{mapleinput}
\mapleresult
\begin{maplelatex}
\[
Kbasis := [[{\displaystyle \frac {1}{2}} \,Id + {\displaystyle \frac
{1}{2}} \,e14, \,
{\displaystyle \frac {1}{2}} \,e1 + {\displaystyle \frac {1}{2}}
\,e4], \,             [Id, \,e1], \,left]
\]
\end{maplelatex}
\end{maplegroup}
\begin{maplegroup}
\begin{mapleinput}
\mapleinline{active}{1d}{cmulQ(f,f); #f is an idempotent in Cl(1,3)}{%}
\end{mapleinput}
\mapleresult
\begin{maplelatex}
\[
{\displaystyle \frac {1}{2}} \,Id + {\displaystyle \frac {1}{2}} \,e14
\]
\end{maplelatex}
\end{maplegroup}
Notice that the generators of the f\/irst list in {\tt 'Kbasis'} are listed
in {\tt Kbasis[2]}. Furthermore, a spinor basis in $S$ over ${\mathbb
K}$ consists of the following two polynomials $f_1$ and~$f_2$:
\begin{maplegroup}
\begin{mapleinput}
\mapleinline{active}{1d}{for i from 1 to nops(Kbasis[1]) do f.i:=Kbasis[1][i] od;}{%}
\end{mapleinput}
\end{maplegroup}
\begin{equation}
f1 := {\displaystyle \frac {1}{2}} \,Id + {\displaystyle \frac {1}{2}} \,e14,
\qquad
f2 := {\displaystyle \frac {1}{2}} \,e1 + {\displaystyle \frac {1}{2}}
\,e4
\label{ablamowicz:eq:fsincl13}
\end{equation}
Using the procedure {\tt 'matKrepr'} we can now f\/ind matrices $m[i]$
with entries in ${\mathbb K}$ representing basis monomials in $\cl_{1,3}.$
Below we will display only matrices representing the $1$-vectors
${\bf e}_1,{\bf e}_2,{\bf e}_3$ and ${\bf e}_4$:
\begin{maplegroup}
\begin{mapleinput}
\mapleinline{active}{1d}{for i from 1 to nops(clibas) do }{%}
\mapleinline{active}{1d}{lprint (`The basis element`,clibas[i],\newline
\MBS  \qquad \qquad `is represented by the following matrix:`);}{%}
\mapleinline{active}{1d}{m[i]:=subs(Id=1,matKrepr(clibas[i])) od;}{%}
\end{mapleinput}
\end{maplegroup}
\mapleresult
\begin{maplettyout}
The basis element   e1   is represented by the following matrix:
\end{maplettyout}
\begin{maplelatex}
\[
{m_{2}} :=  \left[
{\begin{array}{rr}
0 & 1 \\
1 & 0
\end{array}}
 \right]
\]
\end{maplelatex}
\begin{maplettyout}
The basis element   e2   is represented by the following matrix:
\end{maplettyout}
\begin{maplelatex}
\[
{m_{3}} :=  \left[
{\begin{array}{cc}
e2 & 0 \\
0 &  - e2
\end{array}}
 \right]
\]
\end{maplelatex}
\begin{maplettyout}
The basis element   e3   is represented by the following matrix:
\end{maplettyout}
\begin{maplelatex}
\[
{m_{4}} :=  \left[
{\begin{array}{cc}
e3 & 0 \\
0 &  - e3
\end{array}}
 \right]
\]
\end{maplelatex}
\begin{maplettyout}
The basis element   e4   is represented by the following matrix:
\end{maplettyout}
\begin{maplelatex}
\[
{m_{5}} :=  \left[
{\begin{array}{rr}
0 & -1 \\
1 & 0
\end{array}}
 \right]
\]
\end{maplelatex}

Let's def\/ine a $2 \times 2$ quaternionic matrix $A.$ In Maple, we
will represent the standard quaternionic basis $\{1,{\bf i},{\bf
j},{\bf k}\}$ as
{\tt \{1,'ii','jj','kk'\}}.  Later we will make substitutions:
$\mbox{{\tt 'ii'}} \rightarrow e2, \mbox{{\tt 'jj'}} \rightarrow e3,
\mbox{{\tt 'kk'}} \rightarrow e2we3$ since, as we may recall from
Example 3 above,
${\mathbb K}=\{ 1,{\bf e}_2,{\bf e}_3,{\bf e}_{23} \}_{\mathbb R}.$
\begin{maplegroup}
\begin{mapleinput}
\mapleinline{active}{1d}{A:=linalg[matrix](2,2,[1+2*'ii'-3*'kk',2+'ii'
-2*'jj',}{%}
\mapleinline{active}{1d}{'kk'-3*'ii',2*'kk'-2*'jj']); #defining a
quaternionic matrix A}{%}
\end{mapleinput}
\end{maplegroup}
\begin{equation}
A :=  \left[
{\begin{array}{cc}
1 + 2\,ii - 3\,kk & 2 + ii - 2\,jj \\
kk - 3\,ii & 2\,kk - 2\,jj
\end{array}}
 \right]
\label{ablamowicz:eq:Aincl13}
\end{equation}
The isomorphism $\varphi: {\mathbb H}(2) \rightarrow \cl_{1,3}$
has been def\/ined
in Maple through the procedure {\tt 'phi'} (see the Appendix). This
way we can f\/ind image $p$ in $\cl_{1,3}$ of any matrix $A.$ Recall
that {\tt 'FBgens'} in (\ref{ablamowicz:eq:FBgensincl30}) contains the basis
elements of the f\/ield ${\mathbb K}.$
\begin{maplegroup}
\begin{mapleinput}
\mapleinline{active}{1d}{p:=phi(A,m,FBgens);#finding image of A in
Cl(1,3)}{%}
\end{mapleinput}
\mapleresult
\begin{maplelatex}
\begin{eqnarray*}
\lefteqn{p := {\displaystyle \frac {1}{2}} \,Id + e1 + e2 + e3 - e4 -
2\,e12 + e13 +
              {\displaystyle \frac {1}{2}} \,e14 -
              {\displaystyle \frac {1}{2}} \,e23 + e24 + e34 + }\\
 & & {\displaystyle \frac {1}{2}} \,e123 - e124 + e134 +
              {\displaystyle \frac {1}{2}} \,e234 -
              {\displaystyle \frac {5}{2}} \,e1234
%\mbox{\hspace{142pt}}
\end{eqnarray*}
\end{maplelatex}
\end{maplegroup}
The minimal polynomial $p(x)$ of $p$ in $\cl_{1,3}$ is then found
with the procedure {\tt 'climinpoly'}:

\begin{maplegroup}
\begin{mapleinput}
\mapleinline{active}{1d}{climinpoly(p);}{%}
\end{mapleinput}
\mapleresult
\begin{maplelatex}
\[
x^{4} - 2\,x^{3} + 16\,x^{2} + 10\,x + 330
\]
\end{maplelatex}
\end{maplegroup}

So far we have found a Clif\/ford polynomial $p$ in $\cl_{1,3}$ which
is the isomorphic image of the quaternionic matrix $A.$ We will now
compute a sequence of f\/inite power expansions of $p$ using the
procedure {\tt 'sexp'}.  This sequence of Clif\/ford polynomials will
be shown to converge to a polynomial $p_{lim}$ that is the image of
$\exp(A).$ For example, polynomial {\tt p20 = sexp(p,20)} looks as
follows:
\begin{maplegroup}
\begin{mapleinput}
\mapleinline{active}{1d}{for i from 1 to 20 do p.i:=sexp(p,i) od;}{%}
\end{mapleinput}
\mapleresult
\begin{maplelatex}
\begin{eqnarray*}
\lefteqn{p20 := - {\displaystyle \frac
{68240889697169513}{10861169679360000}} \,Id -
{\displaystyle \frac {50515123107772493}{9503523469440000}} \,e34
}\\[1.0ex]
& & \mbox{} + {\displaystyle \frac
{976049744897473}{638892334080000}} \,e123 -
{\displaystyle \frac {76665127748453}{66691392768000}} \,e234
\\[1.0ex]
& & \mbox{} + {\displaystyle \frac
{23336382714907219}{152056375511040000}} \,e124 -
{\displaystyle \frac {1736342897976643}{1974758123520000}} \,e134
\\[1.0ex]
& & \mbox{} + {\displaystyle \frac
{9030311044661089}{1407929402880000}} \,e1234 +
{\displaystyle \frac {802551523836832291}{152056375511040000}} \,e12
\\[1.0ex]
& & \mbox{} - {\displaystyle \frac
{907882088300711}{365520133440000}} \,e13 +
{\displaystyle \frac {4304638284278411}{4472246338560000}} \,e23
\\[1.0ex] & & \mbox{} - {\displaystyle \frac
{360072975386539}{116162242560000}} \,e24 -
{\displaystyle \frac {19812017405738017}{76028187755520000}} \,e14
\\[1.0ex]
& & \mbox{} - {\displaystyle \frac
{1889118161676113}{703964701440000}} \,e1 - {\displaystyle \frac
{277471312336316837}{152056375511040000}} \,e2 \\[1.0ex]
& & \mbox{} - {\displaystyle \frac
{98120514192871531}{152056375511040000}} \,e3 -
{\displaystyle \frac {25277099300039}{44722463385600}}
\,e4
%\mbox{\hspace{100pt}}
\end{eqnarray*}
\end{maplelatex}
\end{maplegroup}

Thus, we have a f\/inite sequence of Clif\/ford polynomials $p_i$
approximating $\exp(p).$ Next, for each of the $16$ basis monomials
present in all polynomials, we create a sequence $s_j$ (or {\tt sj}
in Maple) of its coef\/f\/icients.
\begin{maplegroup}
\begin{mapleinput}
\mapleinline{active}{1d}{for j from 1 to nops(clibas) do}{%}
\mapleinline{active}{1d}{\qquad
s.j:=map(evalf,[seq(coeff(p.i,clibas[j]),i=1..N)]) od:}{%}
\end{mapleinput}
\end{maplegroup}
For example, the sequence {\tt s1} of the coef\/f\/icients of the
identity element $Id$~is:
\begin{maplegroup}
\begin{mapleinput}
\mapleinline{active}{1d}{s1;}{%}
\end{mapleinput}
\mapleresult
\begin{maplelatex}
\begin{eqnarray*}
\lefteqn{[1.500000000, \,-2., \,-6.916666667, \,-18.66666667,
\,-20.22500000, \,-10.85972222, } \\
& & -5.099206349, \,-3.980456349, \,-5.027722663, \,-6.129274691,
\,-6.428549232, \,\\
& & -6.368049418, \,-6.301487892, \,-6.280796253, \,-6.280315663,
\,-6.282290205, \,\\
 & & -6.282986035, \,-6.283054064, \,-6.283026981, \,-6.283014787]
\end{eqnarray*}
\end{maplelatex}
\end{maplegroup}

Having computed the f\/inite sequence of polynomials
$p_1,p_2,\ldots,p_{20},$ one can again verify that this is a
convergent sequence by using any of the Maple's built-in polynomial
norm functions to estimate norms of the dif\/ferences $p_i - p_j$ for
$i, j = 1,\ldots,20.$ It can be again observed that $|p_i - p_j|
\rightarrow 0$ as $i,j \rightarrow \infty.$
Finally, we map back $p_{lim} \simeq p_{20}$ into a $2 \times 2$
matrix {\tt 'expA'} which approximates $\exp (A)$ up to and including
terms of order $N=20.$ After expressing back the basis elements
$\{Id,e2,e3,e2we3\}$ in terms of {\tt \{1,\,'ii',\,'jj',\,'kk'\}} we
obtain:
\begin{maplegroup}
\begin{mapleinput}
\mapleinline{active}{1d}{p_lim:=p20:}{%}
\mapleinline{active}{1d}{expA:=0:for i from 1 to nops(clibas) do}{%}
\mapleinline{active}{1d}{\qquad
expA:=evalm(expA+coeff(p_limit,clibas[i])*m[i]) od:}{%}
\mapleinline{active}{1d}{sexpA:=subs(\{e2we3='kk',e3='jj',e2='ii'\},
evalm(expA));}{%}
\end{mapleinput}
\mapleresult
\begin{maplelatex}
\begin{eqnarray*}
\lefteqn{sexpA := } \\
 & &  \left[ {\vrule height0.80em width0em depth0.80em}
 \right. \!  \!
- {\displaystyle \frac {58889470322671}{8999548740000}}
- {\displaystyle \frac {301630543173}{152472320000}} \,ii
+ {\displaystyle \frac {1778894447566499}{7602818775552000}} \,jj \\
 & & \mbox{}
+ {\displaystyle \frac {560815647244431793}{76028187755520000}} \,kk,
- {\displaystyle \frac {10065855790684619}{4751761734720000}}
- {\displaystyle \frac {5520266650930879}{2534272925184000}} \,ii \\
 & & \mbox{}
+ {\displaystyle \frac {748687448521121}{95995186560000}} \,jj
+ {\displaystyle \frac {203548165276035707}{76028187755520000}} \,kk
\! \! \left.
{\vrule height0.80em width0em depth0.80em} \right]  \\
 & &  \left[ {\vrule height0.80em width0em depth0.80em}\right. \!  \!
- {\displaystyle \frac {30874478783885813}{9503523469440000}}
+ {\displaystyle \frac {33523343384679259}{4001483566080000}} \,ii
+ {\displaystyle \frac {3844312687422001}{1357646209920000}} \,jj \\
 & & \mbox{}
+ {\displaystyle \frac {2613788546323897}{6911653432320000}} \,kk,
- {\displaystyle \frac {228937105237224287}{38014093877760000}}
+ {\displaystyle \frac {127067464810704809}{76028187755520000}} \,ii \\
 & & \mbox{}
+ {\displaystyle \frac {38636486222845507}{25342729251840000}} \,jj
- {\displaystyle \frac {414457945578965819}{76028187755520000}} \,kk
\! \! \left.
{\vrule height0.80em width0em depth0.80em} \right]
\end{eqnarray*}
\end{maplelatex}
\end{maplegroup}
\begin{maplegroup}
\begin{mapleinput}
\mapleinline{active}{1d}{fexpA:=map(evalf,evalm(sexpA));
#floating-point approximation}{%}
\end{mapleinput}
\mapleresult
\begin{maplelatex}
\begin{eqnarray*}
\lefteqn{fexpA:= }\\
& & [ - 6.543602577 - 1.978264272\,ii + .2339782783\,jj +
7.376417403\,kk\,, \\
 & &  - 2.118341860 - 2.178244733\,ii + 7.799218642\,jj + 2.677272355\,kk] \\
& & [ - 3.248740205 + 8.377728618\,ii + 2.831601237\,jj +
.3781712396\,kk\,, \\
 & &  - 6.022426997 + 1.671320448\,ii + 1.524559010\,jj - 5.451372153\,kk
]\mbox{\hspace{6pt}}
\end{eqnarray*}
\end{maplelatex}
\end{maplegroup}
Thus, matrix {\tt 'sexpA'} is the exponential of the quaternionic matrix
$A$ from
(\ref{ablamowicz:eq:Aincl13}) computed with the Clif\/ford algebra $\cl_{1,3}.$

\section{Conclusions}

We have translated the problem of matrix exponentiation ${\mathrm e}^A,\, A
\in {\mathbb K}(n),$ into the problem of computing ${\mathrm e}^p$ in
the Clif\/ford algebra $\cl(Q)$ isomorphic to ${\mathbb K}(n).$
This approach, alternative
to the standard linear algebra methods, is based on the spinor
representation of $\cl(Q)$. It should be equally applicable to other
functions representable as power series. Another use for the
isomorphism between $\cl_{p,q}$ and appropriate matrix rings could be
to f\/inding the Jordan canonical form of $A$ in terms of idempotent
and nilpotent Clif\/ford polynomials from $\cl(Q)$ (see also
\cite{ablamowicz:Sobczyk97} and \cite{ablamowicz:Sobczyk97b} for more on the Jordan form
and its relation to the Clif\/ford algebra). Generally speaking, any
linear algebra property of $A$ can be related to a corresponding
property of $p,$ its isomorphic image in $\cl(Q),$ and it can be
stated in the purely symbolic non-matrix language of the Clif\/ford
algebra. These investigations are greatly facilitated with {\sc
`CLIFFORD'}. At \cite{ablamowicz:Abl97} interested Reader my f\/ind complete Maple
worksheets with the above and other computations.

\section{Acknowledgements}

The author thanks Prof. Thomas McDonald, Department of Mathematics,
Gannon University, Erie, PA, for a critical reading of
this paper, and for bringing to the author's attention a way of
f\/inding the exponential ${\mathrm e}^{At}$ that involves solving a system of
dif\/ferential equations with the Laplace transform method.

\section{Appendix}

The procedures described in this Appendix will work provided the
Maple package {\sc `CLIFFORD'} has been loaded f\/irst into a
worksheet.\footnote[5]{To download {\sc `CLIFFORD'}, see the Web site in
\cite{ablamowicz:Abl97}.} Procedure {\tt 'phi'} was used above to provide the
isomorphism $\varphi$ between the matrix algebras ${\mathbb R}(4),$
${\mathbb C}(2),$
and ${\mathbb H}(2)$ and, respectively, the Clif\/ford algebras $\cl_{3,1},$
$\cl_{3,0},$ and $\cl_{1,3}.$
\begin{maplegroup}
\begin{mapleinput}
\mapleinline{active}{1d}{phi:=proc(A::matrix,m::table,FBgens::list(climon))
\newline
\MBS local N,n,cb,fb,AA,M,a,j,L,sys,vars,sol,p;global B;
   \newline
\MBS if nops(FBgens)=1 then AA:=evalm(A) elif
       \newline
\MBS \quad nops(FBgens)=2 then fb:=op(remove(has,FBgens,Id));
    \newline
\MBS \quad AA:=subs(I=fb,evalm(A)) elif
    \newline
\MBS \quad nops(FBgens)=4 then
fb:=sort(remove(has,FBgens,Id),bygrade); \newline
\MBS \quad AA:=subs({'ii'=fb[1],'jj'=fb[2],'kk'=fb[3]},evalm(A))
\newline
\MBS \quad else ERROR(`wrong number of elements 'FBgens'`) fi;
  \newline
\MBS N:=nops([indices(m)]);n:=linalg[coldim](B):cb:=cbasis(n);
         \newline
\MBS M:=map(displayid,evalm(AA-add(a[j]*m[j],j=1..N)));
       \newline
\MBS L:=map(clicollect,convert(M,mlist));
     \newline
\MBS sys:={op(map(coeffs,L,FBgens))};vars:={seq(a[j],j=1..N)};
         \newline
\MBS sol:=solve(sys,vars); vars:=seq(a[j]*cb[j],j=1..N);
         \newline
\MBS p:=subs(sol,p);RETURN(p)
     \newline
\MBS end:}{%}
\end{mapleinput}
\end{maplegroup}
Procedure {\tt 'climinpoly'} f\/inds a {\it real\/} minimal polynomial
of any Clif\/ford polynomial $p$ in an arbitrary Clif\/ford algebra
$\cl_{p,q}.$
\begin{maplegroup}
\begin{mapleinput}
\mapleinline{active}{1d}{climinpoly:=proc(p::clipolynom,s::string)
\newline
\MBS local dp,L,flag,pp,expr,a,k,eq,sys,vars,sol,poly; \newline
\MBS option remember; \newline
\MBS dp:=displayid(p):L:=[Id,dp];flag:=false:
   \newline
\MBS while not flag do \newline
\MBS \quad pp:=cmul(L[nops(L)],dp): \newline
\MBS \quad expr:=expand(add(a[k]*L[k],k=1..nops(L)));
 \newline
\MBS \quad eq:=clicollect(pp-expr); sys:=coeffs(eq,cliterms(eq));
\newline
\MBS \quad vars:={seq(a[k],k=1..nops(L))}; sol:={solve(sys,vars)}:
\newline
\MBS \quad if sol<>{} then flag:=true else L:=[op(L),pp] fi;
 \newline
\MBS \quad od; \newline
\MBS poly:='x'^nops(L)-add(a[k]*'x'^(k-1),k=1..nops(L));
 \newline
\MBS if nargs=1 then RETURN(sort(subs(op(sol),poly))) \newline
\MBS \quad else RETURN([sort(subs(op(sol),poly)),L]) fi; \newline
\MBS end:}{%}
\end{mapleinput}
\end{maplegroup}
Procedure {\tt 'sexp'} f\/inds a f\/inite formal power series expansion
$\sum\limits_{k=0}^{n}(p^k/k!)$ of any Clif\/ford polynomial $p$ up to and
including the degree specif\/ied as its second argument.  Computation
of the powers of $p$ in $\cl_{p,q}$ is performed modulo the real
minimal polynomial of~$p.$
\begin{maplegroup}
\begin{mapleinput}
\mapleinline{active}{1d}{
   sexp:=proc(p::clipolynom,n::posint) local i,d,L,Lp,pol,poly,k; \newline
\MBS pol:=climinpoly(p,'s');readlib(powmod); \newline
\MBS poly:=add(powmod('x',k,pol[1],'x')/k!,k=0..n); \newline
\MBS L:=[op(poly)];Lp:=[]: \newline
\MBS for i from 1 to nops(L) do \newline
\MBS \quad d:=degree(L[i]); \newline
\MBS \quad if d=0 then Lp:=[op(Lp),L[i]*Id] else \newline
\MBS \quad \quad Lp:=[op(Lp),coeffs(L[i])*pol[2][d+1]] fi od; \newline
\MBS RETURN(add(Lp[i],i=1..nops(Lp))) \newline
\MBS end:}{%}
\end{mapleinput}
\end{maplegroup}

\label{ablamowicz-lp}

\end{document}